# KEY GENERATION AND CERTIFICATION USING MULTILAYER PERCEPTRON IN WIRELESS COMMUNICATION (KGCMLP)

Arindam Sarkar[1] and J. K. Mandal[2]

[1]Department of Computer Science & Engineering, University of Kalyani, W.B, India
`arindam.vb@gmail.com`
[2]Department of Computer Science & Engineering, University of Kalyani, W.B, India
`jkm.cse@gmail.com`

## ABSTRACT

*In this paper, a key generation and certification technique using multilayer perceptron (KGCMLP) has been proposed in wireless communication of data/information. In this proposed KGCMLP technique both sender and receiver uses an identical multilayer perceptrons. Both perceptrons are start synchronization by exchanging some control frames. During synchronization process message integrity test and synchronization test has been carried out. Only the synchronization test does not guarantee the security for this reason key certification phase also been introduced in KGCMLP technique. After Key generation and certification procedure synchronized identical weight vector forms the key for encryption/decryption. Parametric tests have been done and results are compared with some existing classical techniques, which show comparable results for the proposed technique.*

## 1. INTRODUCTION

Key generation and certification is the most significant issue in cryptographic technique. In recent times wide ranges of techniques are developed to protect data and information from eavesdroppers [1, 2, 3, 4, 5, 6, 7, 8, 9]. These algorithms have their virtue and shortcomings. For Example in DES, AES algorithms [1] the cipher block length is nonflexible. In NSKTE [4], NWSKE [5], AGKNE [6], ANNRPMS [7] and ANNRBLC [8] technique uses two neural network one for sender and another for receiver having one hidden layer for producing synchronized weight vector for key generation. Now attacker can get an idea about sender and receiver's neural machine because for each session architecture of neural machine is static. In NNSKECC algorithm [9] any intermediate blocks throughout its cycle taken as the encrypted block and this number of iterations acts as secret key. Here if n number of iterations are needed for cycle formation and if intermediate block is chosen as an encrypted block after $n/2^{th}$ iteration then exactly same number of iterations i.e. n/2 are needed for decode the block which makes easier the attackers life. In this paper KGCMLP technique has been proposed to manage the key generation procedure by synchronizing two multilayer perceptrons and also performs authentication procedure and certified both sender and receiver before communication of actual data.

The organization of this paper is as follows. Section 2 of the paper deals with structure of multilayer perceptron. Proposed key generation and certification technique in KGCMLP has been discussed in section 3. Complexity analysis of the technique is given in section 4. Experimental results are described in section 5. Analysis of the results presented in section 6. Analysis regarding various aspects of the technique has been presented in section 7. Conclusions and future scope are drawn in section 8 and that of references at end.

## 2. STRUCTURE OF MULTILAYER PERCEPTRON

In multilayer perceptron synchronization scheme secret session key is not physically get exchanged over public insecure channel. At end of neural weight synchronization strategy of both parties' generates identical weight vectors and activated hidden layer outputs for both the parties become identical. This identical output of hidden layer for both parties can be use as one time secret session key for secured data exchange. A multilayer perceptron synaptic simulated weight based undisclosed key generation is carried out between recipient and sender. Figure1 shows multilayer perceptron based synaptic simulation system. Sender and receivers multilayer perceptron select same single hidden layer among multiple hidden layers for a particular session. For that session all other hidden layers goes in deactivated mode means hidden (processing) units of other layers do nothing with the incoming input. Either synchronized identical weight vector of sender and receivers' input layer, activated hidden layer and output layer becomes session key or session key can be form using identical output of hidden units of activated hidden layer. The key generation technique and analysis of the technique using random number of nodes (neurons) and the corresponding algorithm is discussed in the subsections 2.1 to 2.5 in details.

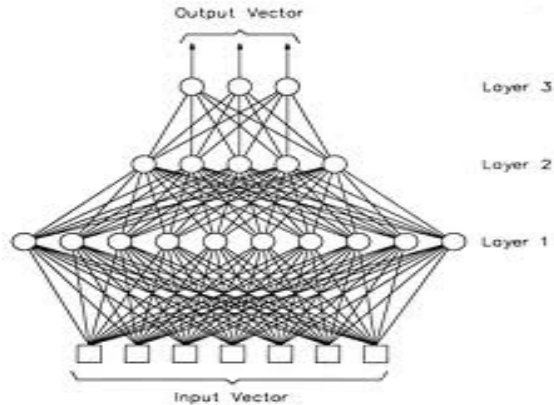

Figure 1. A Multilayer Perceptron with 3 Hidden Layers

Sender and receiver multilayer perceptron in each session acts as a single layer network with dynamically chosen one activated hidden layer and K no. of hidden neurons, N no. of input neurons having binary input vector, $x_{ij} \in \{-1,+1\}$, discrete weights, are generated from input to output, are lies between -L and +L, $w_{ij} \in \{-L,-L+1,...,+L\}$. Where i = 1,…,K denotes the i[th] hidden unit of the perceptron and j = 1,…,N the elements of the vector and one output neuron. Output of the hidden units is calculated by the weighted sum over the current input values . So, the state of the each hidden neurons is expressed using (eq.1)

$$h_i = \frac{1}{\sqrt{N}} w_i x_i = \frac{1}{\sqrt{N}} \sum_{j=1}^{N} w_{i,j} x_{i,j} \qquad (1)$$

Output of the i$^{th}$ hidden unit is defined as

$$\sigma_i = \text{sgn}(h_i) \tag{2}$$

But in case of $h_i = 0$ then $\sigma_i = -1$ to produce a binary output. Hence a, $\sigma_i = +1$, if the weighted sum over its inputs is positive, or else it is inactive, $\sigma_i = -1$. The total output of a perceptron is the product of the hidden units expressed in (eq. 2)

$$\tau = \prod_{i=1}^{K} \sigma_i \tag{3}$$

The learning mechanism proceeds as follows ([6, 7]):

1. If the output bits are different, $\tau^A \neq \tau^B$, nothing is changed.

2. 2. If $\tau^A = \tau^B = \tau$, only the weights of the hidden units with $\sigma_k^{A/B} = \tau^{A/B}$ will be updated.

3. The weight vector of this hidden unit is adjusted using any of the following learning rules:

Anti-Hebbian:

$$W_k^{A/B} = W_k^{A/B} - \tau^{A/B} x_k \Theta(\sigma_k \tau^{A/B})(\tau^A \tau^B) \tag{4}$$

Hebbian;

$$W_k^{A/B} = W_k^{A/B} + \tau^{A/B} x_k \Theta(\sigma_k \tau^{A/B})(\tau^A \tau^B) \tag{5}$$

Random walk

$$W_k^{A/B} = W_k^{A/B} + x_k \Theta(\sigma_k \tau^{A/B})(\tau^A \tau^B) \tag{6}$$

During step (2), if there is at least one common hidden unit with $\sigma_k = \tau$ in the two networks, then there are 3 possibilities that characterize the behaviour of the hidden nodes:

*1. An attractive move:* if hidden units at similar *k* positions have equal output bits,

$\sigma_k^A = \sigma_k^B = \tau^{A/B}$

*2. A repulsive move:* if hidden units at similar *k* positions have unequal output bits, $\sigma_k^A \neq \sigma_k^B$

*3. No move:* when $\sigma_k^A = \sigma_k^B \neq \tau^{A/B}$

The distance between hidden units can be defined by their mutual overlap, $\rho_k$,

$$\rho_k = \frac{w_k^A w_k^B}{\sqrt{w_k^A w_k^A}\sqrt{w_k^B w_k^B}} \tag{7}$$

where $0 < \rho_k < 1$, with $\rho_k = 0$ at the start of learning and $\rho_k = 1$ when synchronization occurs with the two hidden units having a common weight vector.

## 2.1 Multilayer Perceptron Simulation Algorithm

**Input: -** Random weights, input vectors for both multilayer perceptrons.

**Output: -** Secret key through synchronization of input and output neurons as vectors.

**Method:-**

**Step 1.** *Initialization of random weight values of synaptic links between input layer and randomly selected activated hidden layer.*

$$\text{Where, } w_{ij} \in \{-L, -L+1, \ldots +L\} \tag{8}$$

**Step 2.** *Repeat step 3 to 6 until the full synchronization is achieved, using Hebbian-learning rules.*

$$w_{i,j}^+ = g\left(w_{i,j} + x_{i,j}\tau\Theta(\sigma_i\tau)\Theta(\tau^A\tau^B)\right) \tag{9}$$

**Step 3.** *Generate random input vector X. Inputs are generated by a third party or one of the communicating parties.*

**Step 4.** *Compute the values of the activated hidden neurons of activated hidden layer using (eq. 10)*

$$h_i = \frac{1}{\sqrt{N}} w_i x_i = \frac{1}{\sqrt{N}} \sum_{j=1}^{N} w_{i,j} x_{i,j} \tag{10}$$

**Step 5.** *Compute the value of the output neuron using*

$$\tau = \prod_{i=1}^{K} \sigma_i \tag{11}$$

*Compare the output values of both multilayer perceptron by exchanging the system outputs.*

*if Output (A) ≠ Output (B), Go to step 3*
*else if Output (A) = Output (B) then one of the suitable learning rule is applied*
*only the hidden units are trained which have an output bit identical to the common output.*

Update the weights only if the final output values of the perceptron are equivalent. When synchronization is finally achieved, the synaptic weights are identical for both the system.

## 2.2 Multilayer Perceptron Learning rule

At the beginning of the synchronization process multilayer perceptron of A and B start with uncorrelated weight vectors $w_i^{A/B}$. For each time step K, public input vectors are generated randomly and the corresponding output bits $\tau^{A/B}$ are calculated. Afterwards A and B communicate their output bits to each other. If they disagree, $\tau^A \neq \tau^B$, the weights are not changed. Otherwise learning rules suitable for synchronization is applied. In the case of the Hebbian learning rule [10] both neural networks learn from each other.

$$w_{i,j}^+ = g\left(w_{i,j} + x_{i,j}\Theta(\sigma_i\tau)\Theta(\tau^A\tau^B)\right) \tag{12}$$

The learning rules used for synchronizing multilayer perceptron share a common structure. That is why they can be described by a single (eq. 4)

$$w_{i,j}^+ = g\left(w_{i,j} + f(\sigma_i,\tau^A,\tau^B)x_{i,j}\right) \tag{13}$$

with a function $f(\sigma_i,\tau^A,\tau^B)$, which can take the values -1, 0, or +1. In the case of bidirectional interaction it is given by

$$f(\sigma_i,\tau^A,\tau^B) = \Theta(\sigma\tau^A)\Theta(\tau^A\tau^B)\begin{cases}\sigma & \text{Hebbian learning} \\ -\sigma & \text{anti-Hebbian learning} \\ 1 & \text{Random walk learning}\end{cases} \tag{14}$$

The common part $\Theta(\sigma\tau^A)\Theta(\tau^A\tau^B)$ of $f(\sigma_i,\tau^A,\tau^B)$ controls, when the weight vector of a hidden unit is adjusted. Because it is responsible for the occurrence of attractive and repulsive steps [6].

The equation consists of two parts:

1. $\Theta(\sigma\tau^A)\Theta(\tau^A\tau^B)$: This part is common between the three learning rules and it is responsible for the attractive and repulsive effect and controls when the weight vectors of a hidden unit is updated. Therefore, all three learning rules have similar effect on the overlap.

2. $(\sigma, -\sigma, 1)$ : This part differs among the three learning rules and it is responsible for the direction of the weights movement in the space. Therefore, it changes the distribution of the weights in the case of Hebbian and anti-Hebbian learning. For the Hebbian rule, A's ad B's multilayer perceptron learn their own output and the weights are pushed towards the boundaries at −L and +L. In contrast, by using the anti- Hebbian rule, A's and B's multilayer perceptron learn the opposite of their own outputs. Consequently, the weights are pulled from the boundaries ±L. The random walk rule is the only rule that does not affect the weight distribution so they stay uniformly distributed. In fact, at large values of N, both Hebbian and anti-Hebbian rules do not affect the weight distribution. Therefore, the proposed algorithm is restricted to use either random walk learning rule or Hebian or anti-Hebbian learning rules only at large values of *N*. The random walk learning rule is chosen since it does not affect the weights distribution regardless of the value of *N*.

## 2.3 Weight Distribution of Multilayer Perceptron

In case of the Hebbian rule (eq. 8), A's and B's multilayer perceptron learn their own output. Therefore the direction in which the weight $w_{i,j}$ moves is determined by the product $\sigma_i x_{i,j}$. As the output $\sigma_i$ is a function of all input values, $x_{i,j}$ and $\sigma_i$ are correlated random variables. Thus the probabilities to observe $\sigma_i x_{i,j} = +1$ or $\sigma_i x_{i,j} = -1$ are not equal, but depend on the value of the corresponding weight $w_{i,j}$ [11, 13, 14, 15, 16].

$$P(\sigma_i x_{i,j} = 1) = \frac{1}{2}\left[1 + erf\left(\frac{w_{i,j}}{\sqrt{NQ_i - w_{i,j}^2}}\right)\right] \tag{15}$$

According to this equation, $\sigma_i x_{i,j} = \text{sgn}(w_{i,j})$ occurs more often than the opposite, $\sigma_i x_{i,j} = -\text{sgn}(w_{i,j})$. Consequently, the Hebbian learning rule pushes the weights towards the boundaries at -L and +L. In order to quantify this effect the stationary probability distribution of the weights for $t \to \infty$ is calculated for the transition probabilities. This leads to [11].

$$P(w_{i,j} = w) = P_0 \prod_{m=1}^{|w|} \frac{1 + erf\left(\frac{m-1}{\sqrt{NQ_i - (m-1)^2}}\right)}{1 - erf\left(\frac{m}{\sqrt{NQ_i - m^2}}\right)} \tag{16}$$

Here the normalization constant $\rho_0$ is given by

$$P_0 = \left(\sum_{w=-L}^{L} \prod_{m=1}^{|w|} \frac{1 + erf\left(\frac{m-1}{\sqrt{NQ_i - (m-1)^2}}\right)}{1 - erf\left(\frac{m}{\sqrt{NQ_i - m^2}}\right)}\right)^{-1} \tag{17}$$

In the limit $N \to \infty$ the argument of the error functions vanishes, so that the weights stay uniformly distributed. In this case the initial length of the weight vectors is not changed by the process of synchronization.

$$\sqrt{Q_i(t=0)} = \sqrt{\frac{L(L+1)}{3}} \tag{18}$$

But, for finite N, the probability distribution itself depends on the order parameter $Q_i$ Therefore its expectation value is given by the solution of the following equation:

$$Q_i = \sum_{w=-L}^{L} w^2 P(w_{i,j} = w) \qquad (19)$$

## 2.4 Order Parameters

In order to describe the correlations between two multilayer perceptron caused by the synchronization process, one can look at the probability distribution of the weight values in each hidden unit. It is given by $(2L + 1)$ variables.

$$P_{a,b}^i = P(w_{i,j}^A = a \wedge w_{i,j}^B = b) \qquad (20)$$

which are defined as the probability to find a weight with $w_{i,j}^A = a$ in A's multilayer perceptron and $w_{i,j}^B = b$ in B's multilayer perceptron. In both cases, simulation and iterative calculation, the standard order parameters, which are also used for the analysis of online learning, can be calculated as functions of $P_{a,b}^i$ [12].

$$Q_i^A = \frac{1}{N} w_i^A w_i^A = \sum_{a=-L}^{L} \sum_{b=-L}^{L} a^2 P_{a,b}^i \qquad (21)$$

$$Q_i^B = \frac{1}{N} w_i^B w_i^B = \sum_{a=-L}^{L} \sum_{b=-L}^{L} b^2 P_{a,b}^i \qquad (22)$$

$$R_i^{AB} = \frac{1}{N} w_i^A w_i^B = \sum_{a=-L}^{L} \sum_{b=-L}^{L} ab P_{a,b}^i \qquad (23)$$

Then the level of synchronization is given by the normalized overlap between two corresponding hidden units

$$\rho_i^{AB} = \frac{w_i^A w_i^B}{\sqrt{w_i^A w_i^A} \sqrt{w_i^B w_i^B}} = \frac{R_i^{AB}}{\sqrt{Q_i^A Q_i^B}} \qquad (24)$$

## 2.5 Hidden Layer as a Secret Session Key

At end of full weight synchronization process, weight vectors between input layer and activated hidden layer of both multilayer perceptron systems become identical. Activated hidden layer's output of source multilayer perceptron is used to construct the secret session key. This session key is not get transmitted over public channel because receiver multilayer perceptron has same identical activated hidden layer's output. Compute the values of the each hidden unit by

$$\sigma_i = \mathrm{sgn}\left(\sum_{j=1}^{N} w_{ij} x_{ij}\right) \quad \mathrm{sgn}(x) = \begin{cases} -1 & \text{if } x < 0, \\ 0 & \text{if } x = 0, \\ 1 & \text{if } x > 0. \end{cases} \qquad (25)$$

For example consider 8 hidden units of activated hidden layer having absolute value (1, 0, 0, 1, 0, 1, 0, 1) becomes an 8 bit block. This 10010101 become a secret session key for a particular session and cascaded xored with recursive replacement encrypted text. Now final session key based encrypted text is transmitted to the receiver end. Receiver has the identical session key i.e. the output of the hidden units of activated hidden layer of receiver. This session key used to get the recursive replacement encrypted text from the final cipher text. In the next session both the machines started tuning again to produce another session key.

Identical weight vector derived from synaptic link between input and activated hidden layer of both multilayer perceptron can also becomes secret session key for a particular session after full weight synchronization is achieved.

## 3. KEY GENERATION AND CERTIFICATION IN KGCMLP

The key organization procedure defines messages and data necessary for cryptographic keys organization with certification. Both receiver and sender use one multilayer perceptron with identical structure. The parameters k, l and n are public. For example if multilayer perceptron uses 3 hidden neurons in hidden layer and each hidden neurons has 32 inputs neurons and weights limit equal ±127, then total number of weight generation is 96. 8 binary bits are needed to represents each weight, where the MSB represents the signal:

$$= \begin{cases} 1 & if\ MSB = 1, \\ 0 & if\ otherwise \end{cases} \quad (26)$$

and others 7 bits represents the absolute value of the weight. In the time of multilayer perceptron synchronizing process, a total of 768 bits (total number of weights is 96 and 8 bits are needed to represents each weight) i.e. 6 groups of 128 bits gets generated. Proposed KGCMLP protocol also crates some control frames, these frames are shown in figure 2. The table 1 shows the frames and their respective command codes.

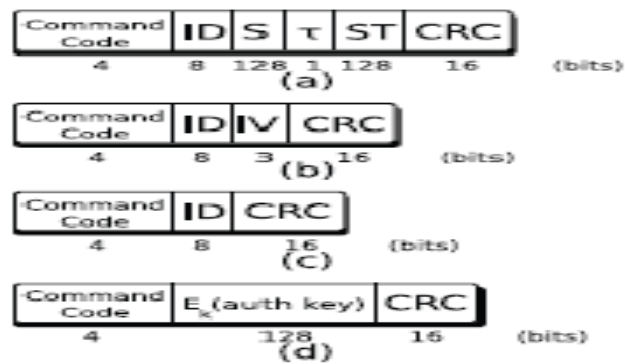

Figure 2. Frames of KGCMLP protocol (a) SYN, (b) FIN_SYN,
(c) ACK_SYN, NAK_ SYN and (d) AUTH.

Table 1. KGCMLP Frames and their command codes.

| Frame | Command |
|---|---|
| SYN | 0000 |
| FIN_SYN | 0001 |
| ACK_SYN | 0010 |
| NAK_SYN | 0011 |
| AUTH | 0100 |
| Reserved | 0101-1111 |

The process of KGCMLP protocol is divided in two phases: keys generation and certification. The key generation phase is shown in figure 3. It begins with the assignment of random values to weights. The input vector (X) is created by the sender at each step, through a seed of 128 bits.

The sender uses the frame ACK_SYN to notify the receiver: seed value (S), its output ($\tau^{sender}$), an encrypted sequence of bits ($E_k(ST)$) and an identifier (ID). The encrypted sequence is obtained encrypting a variable known, called ST. This is necessary for the synchronization test. The identifier is the function of informing the sender and receiver where the message is a recent message. The variable ID starts with zero and is incremented every time that the sender sends a synchronization frame.

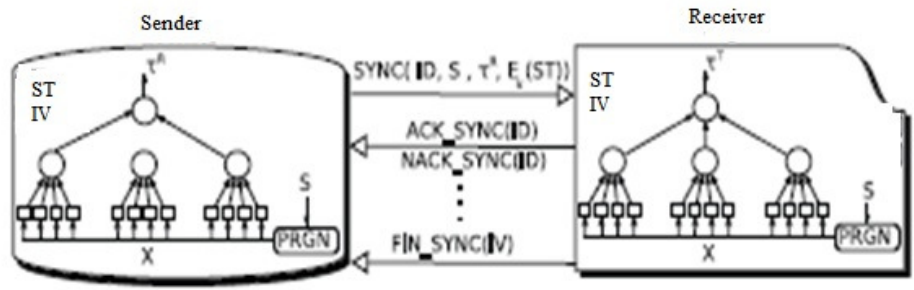

Figure 3. Exchange of messages between sender and receiver during key generation.

After transmitting the frame SYN, the sender starts a timer and waits a reply from the receiver. If the receiver does not take action until a certain limit time and number of attempts has not exceeded a certain value, the sender restarts the synchronization procedure. When the receiver receives the frame SYN, the receiver should carry out integrity test. If the messages are received as sent (with no replication, incorporation, alteration, reordering, or replay) the receiver will execute the synchronization check. This check is accomplished through the following steps: receiver utilize its 128 first weights as key for decryption $E_k(ST)$ that was received from the sender. If the result is identical to ST variable formerly stored in its memory, the networks are synchronized. Finally, the receiver should arbitrarily pick one of six positions of weights vector to form a key. Afterward, it should inform the sender who obtained the synchronization and which will be the IV (IndexVector) of weights that will be used to the generate key. The receiver should send the frame FIN_SYN to alert the sender. If decryption algorithm does not produce predictable result, the receiver should use the seed (S) in its pseudo-random number generator to produce the network inputs (X). With this input vector the receiver will work out its output ($\tau^{receiver}$). If network output is equal to sender output ($\tau^{receiver} = \tau^{sender}$) then receiver should regulate their weights. At the end of weights update, the receiver should report the sender that outputs are the equal. The receiver uses the frame ACK_SYN to notify the sender, with the same ID value received from sender. If the receiver and sender outputs are different, the receiver should not fine-tune its weights and inform the sender its output. The receiver sends the message NACK_SYN to notify the sender, with the same ID value. If sender receives ACK _SYN it should update its weights. The sender will create new synchronization frame until receive the frame FIN_ACK from receiver. When the sender receives the frame FIN_ACK, it must extract the keys of the neural network according to the Index Vector informed by the receiver. At the end of the synchronization, both networks provide the same key for encryption. However, only the process of generating keys does not guarantee the information security. Therefore, any attacker can also synchronize with an authorized device, because the protocol is a public knowledge. Thus, to ensure that only entities authorized have access to information is necessary authentication service. The function of the authentication service is to ensure the recipient that the message is from the source that it claims. There are several authentication methods, differentiated mainly by the use of secret-keys or public-keys. Unlike encryption algorithms, in public-key authentication the user A send message encrypted with A.'s private-key. The recipient of the message uses the public-key to verify the message, thus ensuring that only the owner of the private-key could have encrypted the message. On secret keys authentication both entities must have a common secret code. In this paper two secret codes are used, called RSC (Receiver Secret Code) and SSC (Sender Secret Code), as shown in the figure 4.

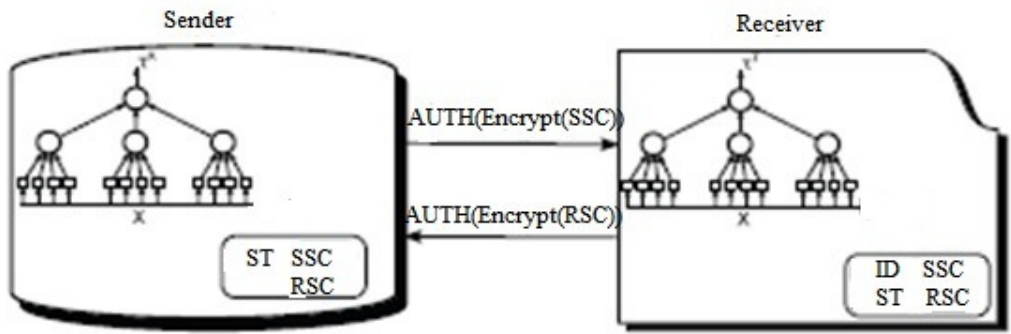

Figure 4. Exchange of messages during key certification.

The key certification start with both multilayer perceptron synchronized. The sender will use the neural network weights as key to encrypt the variable SSC. The sender sends an authentication frame (AUTH) to receiver. If the receiver does not respond until a certain limit time, the sender increases the number of attempts. If this number does not exceed a threshold, the sender sends again the AUTH frame. Otherwise, sender terminates the authentication phase. The receiver should decrypt the key field when the AUTH frame is received. If the result is equal to receiver SSC, the tag learns that the synchronized device is authorized. After that, the receiver must be authenticated. For this, the receiver should use neural network weights to encrypt its RSC variable and send it to sender. The sender receives the AUTH frame, decrypt the key field and verify the RSC. If the received key is valid the receiver is certified, terminates the authentication phase. In data transmission phase, the sender can continue to generate random seeds, to feed the multilayer perceptron, and get different keys for each frame transmitted.

## 4. COMPLEXITY ANALYSIS

The complexity of the Synchronization technique will be O(L), which can be computed using following three steps.

**Step 1.** To generate a MLP guided key of length N needs O(N) Computational steps. The average synchronization time is almost independent of the size N of the networks, at least up to N=1000.Asymptotically one expects an increase like O (log N).

**Step 2.** Complexity of the encryption technique is O(L).

*Step 2. 1.* Recursive replacement of bits using prime nonprime recognition encryption process takes O(L).

*Step 2. 2.* MLP based encryption technique takes O(L) amount of time.

**Step 3.** Complexity of the decryption technique is O(L).

*Step 3. 1.* In MLP based decryption technique, complexity to convert final cipher text into recursive replacement cipher text T takes O(L).

*Step 3. 2.* Transformation of recursive replacement cipher text T into the corresponding stream of bits $S = s_0 \, s_1 \, s_2 \, s_3 \, s_4 \ldots s_{L-1}$, which is the source block takes O(L) as this step also takes constant amount of time for merging $s_0 \, s_1 \, s_2 \, s_3 \, s_4 \ldots s_{L-1}$.

## 5. EXPERIMENT RESULTS

The average of synchronization times with respect to the variable l are shown in figure 5. In the graphic, we can observe that the increase in safety level (i.e., the increase in the variable l) implies in an exponential increase of iterations needed for synchronization.

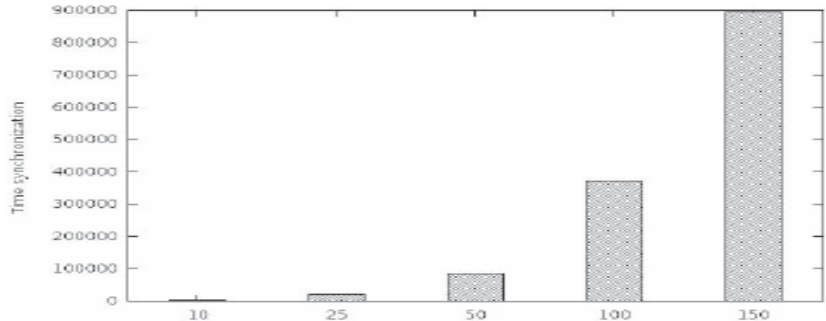

Figure 5. Average of synchronization time as a function of l for k = 3 and n = 32, obtained in 1000 runs.

While the parameter l influence directly in the safety level, the variable n determines the number of generated keys. Figure 6 shows the relationship between the average synchronization times as a function of n in 1000 samples. The graphic shown that the increase of n (i.e., larger number of generated keys) implies a small increase in the average time needed to obtain synchronization.

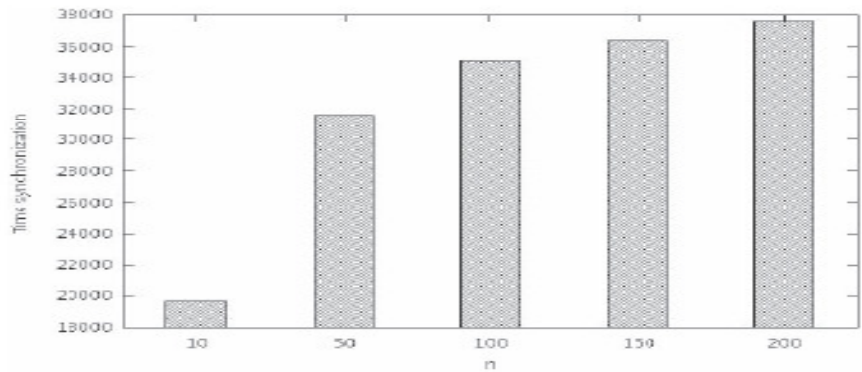

Figure 6. Average of time synchronization as a function of n for k = 3 and l = 5, found from 1000 runs.

In this section the results of implementation of the proposed KGCMLP encryption/decryption technique has been presented in terms of encryption decryption time, Chi-Square test, source file size vs. encryption time along with source file size vs. encrypted file size. The results are also compared with existing RSA [1] technique, existing ANNRBLC [8] and NNSKECC [9].

.Table 2. Encryption / decryption time vs. File size

| Encryption Time (s) | | | Decryption Time (s) | | |
|---|---|---|---|---|---|
| Source Size (bytes) | KGCMLP | NNSKECC [9] | Encrypted Size (bytes) | KGCMLP | NNSKECC [9] |
| 18432 | 6. 42 | 7.85 | 18432 | 6.99 | 7.81 |
| 23044 | 9. 23 | 10.32 | 23040 | 9.27 | 9.92 |
| 35425 | 14. 62 | 15.21 | 35425 | 14. 47 | 14.93 |
| 36242 | 14. 72 | 15.34 | 36242 | 15. 19 | 15.24 |
| 59398 | 25. 11 | 25.49 | 59398 | 24. 34 | 24.95 |

Table 2 shows encryption and decryption time with respect to the source and encrypted size respectively. It is also observed the alternation of the size on encryption.

In figure 7 stream size is represented along X axis and encryption / decryption time is represented along Y-axis. This graph is not linear, because of different time requirement for finding appropriate KGCMLP key. It is observed that the decryption time is almost linear, because there is no KGCMLP key generation process during decryption.

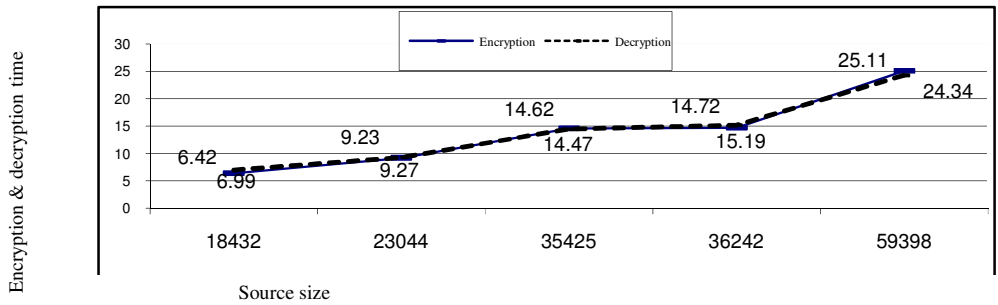

Figure 7. Encryption decryption time against stream size

Table 3 shows Chi-Square value for different source stream size after applying different encryption algorithms. It is seen that the Chi-Square value of KGCMLP is better compared to the algorithm ANNRBLC [8] and comparable to the Chi-Square value of the RSA algorithm.

Table 3. Source size vs. Chi-Square value

| Stream Size (bytes) | Chi-Square value (TDES) [1] | Chi-Square value in (KGCMLP) | Chi-Square value (ANNRBLC) [8] | Chi-Square value (RSA) [1] |
|---|---|---|---|---|
| 1500 | 1228.5803 | 2856.2673 | 2471.0724 | 5623.14 |
| 2500 | 2948.2285 | 6582.7259 | 5645.3462 | 22638.99 |
| 3000 | 3679.0432 | 7125.2364 | 6757.8211 | 12800.355 |
| 3250 | 4228.2119 | 7091.1931 | 6994.6198 | 15097.77 |
| 3500 | 4242.9165 | 12731.7231 | 10572.4673 | 15284.728 |

Figure 8 shows graphical representation of table 3.

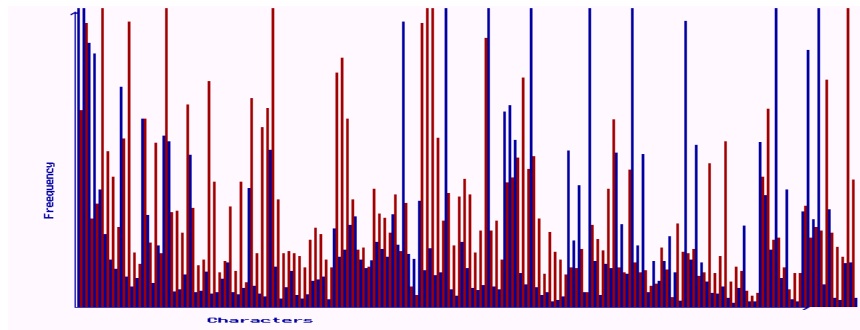

Figure 8. Chi-Square value against stream size

Table 4 shows total number of iteration needed and number of data being transferred for KGCMLP key generation process with different numbers of input(N) and activated hidden(H) neurons and varying synaptic depth(L).

Table 4. Data Exchanged and No. of Iterations For Different Parameters Value

| No. of Input Neurons(N) | No. of Activated Hidden Neurons(K) | Synaptic Weight (L) | Total No. of Iterations | Data Exchanged (Kb) |
|---|---|---|---|---|
| 5 | 15 | 3 | 624 | 48 |
| 30 | 4 | 4 | 848 | 102 |
| 25 | 5 | 3 | 241 | 30 |
| 20 | 10 | 3 | 1390 | 276 |
| 8 | 15 | 4 | 2390 | 289 |

Following figure 9. Shows the snapshot of KGCMLP key simulation process.

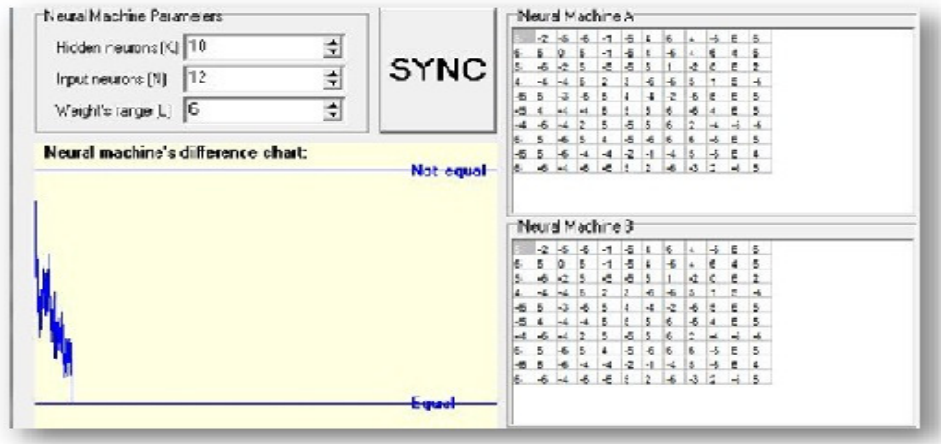

Figure 9. KGCMLP Key Simulation Snapshot with N=12, K=10 and L=6

## 6. ANALYSIS OF RESULTS

From results obtained it is clear that the technique will achieve optimal performances. Encryption time and decryption time varies almost linearly with respect to the block size. For the algorithm presented, Chi-Square value is very high compared to some existing algorithms. A user input key has to transmit over the public channel all the way to the receiver for performing the decryption procedure. So there is a likelihood of attack at the time of key exchange. To defeat this insecure secret key generation technique a neural network based secret key generation technique has been devised. The security issue of existing algorithm can be improved by using KGCMLP secret session key generation technique. In this case, the two partners A and B do not have to share a common secret but use their indistinguishable weights or output of activated hidden layer as a secret key needed for encryption. The fundamental conception of KGCMLP based key exchange protocol focuses mostly on two key attributes of KGCMLP. Firstly, two nodes coupled over a public channel will synchronize even though each individual network exhibits disorganized behaviour. Secondly, an outside network, even if identical to the two communicating networks, will find it exceptionally difficult to synchronize with those parties, those parties are communicating over a public network. An attacker E who knows all the particulars of the algorithm and records through this channel finds it thorny to synchronize with the parties, and hence to calculate the common secret key. Synchronization by mutual learning (A and B) is much quicker than learning by listening (E) [10]. For usual cryptographic systems, we can improve the safety of the protocol by increasing of the key length. In the case of KGCMLP, we improved it by increasing the synaptic depth L of the neural networks. For a brute force attack using K hidden neurons, K*N input neurons and boundary of weights L, gives (2L+1)KN possibilities. For example, the configuration K = 3, L = 3 and N = 100 gives us 3*10253 key possibilities, making the attack unfeasible with today's computer power. E could start from all of the (2L+1)3N initial weight vectors and calculate the ones which are consistent with the input/output sequence. It has been shown, that all of these initial states move towards the same final weight vector, the key is unique. This is not true for simple perceptron the most unbeaten cryptanalysis has two supplementary ingredients first; a group of attacker is used. Second, E makes extra training steps when A and B are quiet [10]-[12]. So increasing synaptic depth L of the KGCMLP we can make our KGCMLP safe.

## 7. SECURITY ISSUE

The main difference between the partners and the attacker in KGCMLP is that A and B are able to influence each other by communicating their output bits $\tau^A$ & $\tau^B$ while E can only listen to these messages. Of course, A and B use their advantage to select suitable input vectors for adjusting the weights which finally leads to different synchronization times for partners and attackers. However, there are more effects, which show that the two-way communication between A and B makes attacking the KGCMLP protocol more difficult than simple learning of examples. These confirm that the security of KGCMLP key generation is based on the bidirectional interaction of the partners. Each partener uses a seperate, but identical pseudo random number generator. As these devices are initialized with a secret seed state shared by A and B. They produce exactly the same sequence of input bits. Whereas attacker does not know this secret seed state. By increasing synaptic depth average synchronize time will be increased by polynomial time. But success probability of attacker will be drop exponentially Synchonization by mutual learning is much faster than learning by adopting to example generated by other network. Unidirectional learning and bidirectional synchronization. As E can't influence A and B at the time they stop transmit due to synchrnization. Only one weight get changed where, = T. So, difficult to find weight $w_i$ for attacker to know the actual weight without knowing internal representation it has to guess.

## 8. FUTURE SCOPE & CONCLUSION

This paper presented a novel approach for key generation and authentication using multilayer perceptron. KGCMLP algorithms is proposed as extensions to the ordinary mutual learning algorithm. This algorithm can be used in many applications such as video and voice conferences. This technique enhances the security features of the key exchange algorithm by increasing of the synaptic depth L of the KGCMLP. Here two partners A and B do not have to exchange a common secret key over a public channel but use their indistinguishable weights or outputs of the activated hidden layer as a secret key needed for encryption or decryption. So likelihood of attack proposed technique is much lesser than the simple key exchange algorithm.

Future scope of this technique is that this KGCMLP model can be used in wireless communication and also in key distribution mechanism.

## ACKNOWLEDGEMENTS

The author expresses deep sense of gratitude to the Department of Science & Technology (DST) , Govt. of India, for financial assistance through INSPIRE Fellowship leading for a PhD work under which this work has been carried out, at the department of Computer Science & Engineering, University of Kalyani.

## REFERENCES

[1] Atul Kahate, Cryptography and Network Security, 2003, Tata McGraw-Hill publishing Company Limited, Eighth reprint 2006.
[2] Sarkar Arindam, Mandal J. K, "Artificial Neural Network Guided Secured Communication Techniques: A Practical Approach" LAP Lambert Academic Publishing ( 2012-06-04 ), ISBN: 978-3-659-11991-0, 2012
[3] Sarkar Arindam, Karforma S, Mandal J. K, "Object Oriented Modeling of IDEA using GA based Efficient Key Generation for E-Governance Security (OOMIG) ", International Journal of Distributed and Parallel Systems (IJDPS) Vol.3, No.2, March 2012, DOI : 10.5121/ijdps.2012.3215, ISSN : 0976 - 9757 [Online] ; 2229 - 3957 [Print]. Indexed by: EBSCO, DOAJ, NASA, Google Scholar, INSPEC and WorldCat, 2011.
[4] Mandal J. K., Sarkar Arindam, "Neural Session Key based Traingularized Encryption for Online Wireless Communication (NSKTE)", 2nd National Conference on Computing and Systems, (NaCCS 2012), March 15-16, 2012, Department of Computer Science, The University of Burdwan, Golapbag North, Burdwan –713104, West Bengal, India. ISBN 978- 93-808131-8-9, 2012.
[5] Mandal J. K., Sarkar Arindam, "Neural Weight Session Key based Encryption for Online Wireless Communication (NWSKE)", Research and Higher Education in Computer Science and Information Technology, (RHECSIT- 2012) ,February 21-22, 2012, Department of Computer Science, Sammilani Mahavidyalaya, Kolkata , West Bengal, India. ISBN 978-81- 923820-0-5,2012
[6] Mandal J. K., Sarkar Arindam, "An Adaptive Genetic Key Based Neural Encryption For Online Wireless Communication (AGKNE)", International Conference on Recent Trends In Information Systems (RETIS 2011) BY IEEE, 21-23 December 2011, Jadavpur University, Kolkata, India. ISBN 978-1-4577-0791-9, 2011
[7] Mandal J. K., Sarkar Arindam, "An Adaptive Neural Network Guided Secret Key Based Encryption Through Recursive Positional Modulo-2 Substitution For Online Wireless Communication (ANNRPMS)", International Conference on Recent Trends In Information Technology (ICRTIT 2011) BY IEEE, 3-5 June 2011, Madras Institute of Technology, Anna University, Chennai, Tamil Nadu, India. 978-1-4577-0590-8/11, 2011
[8] Mandal J. K., Sarkar Arindam, "An Adaptive Neural Network Guided Random Block Length Based Cryptosystem (ANNRBLC)", 2[nd] International Conference on Wireless Communications, Vehicular Technology, Information Theory And Aerospace & Electronic System Technology" (Wireless Vitae 2011) By IEEE Societies, February 28- March 03, 2011,Chennai, Tamil Nadu, India. ISBN 978-87-92329-61-5, 2011
[9] Mandal J. K., Sarkar Arindam, "Neural Network Guided Secret Key based Encryption through Cascading Chaining of Recursive Positional Substitution of Prime Non-Prime (NNSKECC)", International Confference on Computing and Systems, ICCS – 2010,  19–20 November, 2010,Department of Computer Science, The University of Burdwan, Golapbag North, Burdwan – 713104, West Bengal, India.ISBN 93-80813-01-5, 2010
[10] R. Mislovaty, Y. Perchenok, I. Kanter, and W. Kinzel. Secure key-exchange protocol with an absence of injective functions. Phys. Rev. E, 66:066102,2002.
[11] A. Ruttor, W. Kinzel, R. Naeh, and I. Kanter. Genetic attack on neural cryptography. Phys. Rev. E, 73(3):036121, 2006.
[12] A. Engel and C. Van den Broeck. Statistical Mechanics of Learning. Cambridge University Press, Cambridge, 2001.
[13] T. Godhavari, N. R. Alainelu and R. Soundararajan "Cryptography Using Neural Network " IEEE Indicon 2005 Conference, Chennai, India, 11-13 Dec. 2005.
[14] Wolfgang Kinzel and ldo Kanter, "Interacting neural networks and cryptography", Advances in Solid State Physics, Ed. by B. Kramer   (Springer, Berlin. 2002), Vol. 42, p. 383 arXiv- cond-mat/0203011, 2002
[15] Wolfgang Kinzel and ldo Kanter, "Neural cryptography"  proceedings of the 9[th] international conference on Neural    Information processing(ICONIP 02).
[16] Dong Hu "A new service based computing security model with neural cryptography"IEEE07/2009.

## Authors

**Arindam Sarkar**
INSPIRE Fellow (DST, Govt. of India), MCA (VISVA BHARATI, Santiniketan, University First Class First Rank Holder), M.Tech (CSE, K.U, University First Class First Rank Holder). Total number of publications 13.

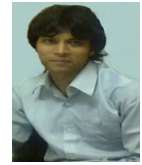

**Jyotsna Kumar Mandal**
M. Tech.(Computer Science, University of Calcutta), Ph.D.(Engg., Jadavpur University) in the field of Data Compression and Error Correction Techniques, Professor in Computer Science and Engineering, University of Kalyani, India. Life Member of Computer Society of India since 1992 and life member of cryptology Research Society of India. Dean Faculty of Engineering, Technology & Management, working in the field of Network Security, Steganography, Remote Sensing & GIS Application, Image Processing. 25 years of teaching and research experiences. Eight Scholars awarded Ph.D. one submitted and 8 are pursuing. Total number of publications 248.

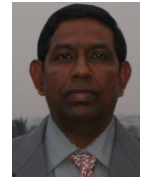